\documentclass[
prl,
aps,
amsmath,
amssymb,
]{revtex4} 
\usepackage{subfigure}
\usepackage{color}
\usepackage{latexsym}
\usepackage{mathrsfs}

\DeclareMathOperator{\Tr}{Tr}

\ifx\pdfoutput\undefined
\usepackage{graphicx}
\else
\usepackage[pdftex]{graphicx}
\usepackage{epstopdf}
\fi 

\begin{document}

\title{Unitarity in $WW \to WW$ elastic scattering in topologically
  massive SU(2) gauge theory}      
\author{Amitabha Lahiri}\email{amitabha@bose.res.in} 
\author{Debmalya Mukhopadhyay}\email{debmalya@bose.res.in}
\affiliation{S. N. Bose National Centre for Basic Sciences, \\ 
Block JD, Sector III, Salt Lake, Calcutta 700 098, INDIA\\ 
}

\begin{abstract}

  We consider the elastic scattering of longitudinally polarized
  gauge bosons in an SU(2) generalization of topologically massive
  gauge theory in four dimensions. We show that the amplitude
  remains finite at large $s$, even though the theory does not
  contain a Higgs particle, in contradiction to common lore.  
\end{abstract}
\date\today

\maketitle


 

The Higgs boson~\cite{Higgs:1964pj,Englert:1964et,Guralnik:1964eu}
provides a mechanism of
mass generation for gauge bosons. It also plays an important role
in maintaining the unitarity of tree-level scattering process among
the massive gauge bosons, $W$. The total amplitude for elastic
scattering of longitudinally polarized massive gauge bosons
diverges with the increase of energy unless the Higgs particle is
included as a mediator~\cite{Llewellyn 
  Smith:1973ey, Cornwall:1973tb, Cornwall:1974km,
  Joglekar:1973hh}. 

 Here we consider the topological mass generation mechanism in
 $3+1$ dimensions, in which the mass of the gauge boson is
 generated without recourse to spontaneous symmetry breaking. The
 mass is generated via an interaction of the form $B\wedge F$,
 where $B$ is an antisymmetric 2-tensor potential and $F$ is the
 field strength of the gauge field~\cite{Allen:1990gb}. The mass
 itself is the coupling coefficient of the term and does not appear
 via any separate mechanism, but is introduced by hand. The
 resulting theory does not have a Higgs like scalar particle, so it
 is necessary to check whether unitarity is violated in tree-level
 processes.    
  

 In this model all gauge bosons have the same mass $m$. The theory
 can be formulated for any gauge group but for our purposes it is
 sufficient to consider the group SU(2). Let us label the gauge
 bosons as $W^+, W^-$ and $W^3$, with 
 \begin{eqnarray}
W^{\pm}=\frac{W^1\mp iW^2}{\sqrt{2}}\,.
\end{eqnarray}
  Note that $W_3$ has the same mass as $W^\pm$
 
\begin{figure}[hbt]
    \includegraphics[scale=0.4]{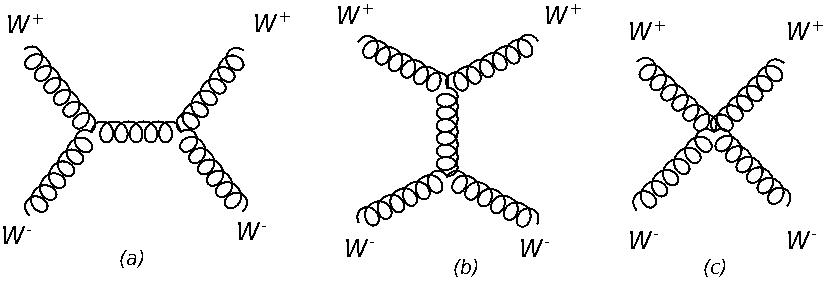} 
\caption{(a) s-channel, (b) t-channel, (c) direct quartic. }
  \label{fig:Wonly}
\end{figure}

 Let us first go over the arguement of how unitarity may be
 violated by a massive triplet of gauge bosons. The tree-level
 scattering processes involving only the $W$ bosons 
are shown in Fig.~\ref{fig:Wonly}.

Writing the magnitude of the 3-momentum in the center of momentum frame
as $P$, and the cosine of the angle between the initial and final
$W^+$ as $c\,,$ we find that in the high energy limit $P \gg m$,
the Feynman amplitudes of these diagrams go
as~\cite{Joglekar:1973hh}
\begin{eqnarray}
{\cal M}_{1s} &=& -4\frac{g^2 P^4}{m^4}c - 9\frac{g^2 P^2}{m^2}c +
O(P^0)\\ 
{\cal M}_{1t} &=& \frac{g^2 P^4}{m^4}(1 - c)(3 + c) + \frac{g^2
  P^2}{2m^2}(9 + 7c - 4c^2) + O(P^0)\\ 
{\cal M}_{1q} &=& -\frac{g^2 P^4}{m^4}(3 - 6c -c^2) - 2 \frac{g^2
  P^2}{m^2} (2 - 3c - c^2) + O(P^0)\\ 
\sum {\cal M}_1 &=& \frac{g^2P^2}{2m^2}(1 + c) + O(P^0)\,.
\end{eqnarray}
Clearly, if these are the only diagrams for the $W_L^+ W_L^-$
elastic scattering process, the amplitude diverges as $P\to \infty$
and unitarity is violated. We note that the sum of
these amplitudes for the SU(2)$\times$U(1) gauge theory, when the
mediating gauge boson can be either a $Z$ or a photon, may be
written in the same form. 

\begin{figure}[hbt]
    \includegraphics[scale=0.4]{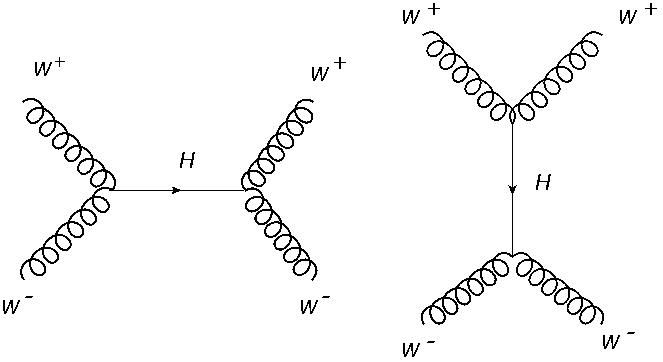} 
\caption{Higgs-mediated diagrams}
  \label{fig:H-mediated}
\end{figure}

Let us see how this divergence is cancelled in the Standard
Model. There the masses of the gauge bosons come from the Higgs
mechanism, i.e. from a complex scalar doublet. Three of the four
modes are converted into the longitudinal modes of the three gauge
bosons, and the fourth remains as the Higgs particle $H$, coupling
to $W^\pm$ via a term $m H W_\mu^+ W^{\mu -}$. This leads to two
additional diagrams with the Higgs particle as the mediator, shown
in Fig.~\ref{fig:H-mediated},
%
%
corresponding to amplitudes
\begin{eqnarray}
{\cal M}_{2s} & = & \frac{g^2 P^2}{2 m^2}(1 - c) + O(P^0)\\
{\cal M}_{2t} & = &	-\frac{g^2 P^2}{m^2}  + O(P^0)\,,
\end{eqnarray}
so that the total amplitude from the Higgs mediated diagrams cancel
the divergence of the previous diagrams, the $W_L^+ W_L^-$ elastic
scattering amplitude remains finite, and unitarity is not violated
as $P\to \infty\,.$ 
A crucial ingredient here was the 3-point coupling between the
Higgs and the $W$ particles.  

For the topological mass generation mechanism these two diagrams do
not appear. However, some new diagrams do appear because of the
presence of new fields and interactions. Let us check if the
scattering process remains unitary. In this model the mass of the
gauge bosons is provided by a coupling of the field with an
antisymmetric tensor field $B_{\mu\nu}$.  We write the Lagrangian
density as
\begin{eqnarray}
\mathscr{L}=-\frac{1}{4}F^{\mu\nu}_a F_{\mu\nu}^a+\frac{1}{12}H^{\mu\nu\lambda}_a H_{\mu\nu\lambda}^a+\frac{m}{4}\epsilon ^{\mu\nu\rho\lambda}F_{\mu\nu}^a B_{\rho\lambda}^a
\end{eqnarray}
 
Here  $F^{\mu\nu}_a$ is the field strength of the gauge bosons
\begin{eqnarray}
F^{\mu\nu}_a&=& \partial^\mu W^\nu_a-\partial^\nu W^\mu_a -
gf_{bca} W^\mu_b W^\nu_c \\ &=&\mathscr{F}^{\mu\nu}_a- gf_{bca}
W^\mu_b W^\nu_c  
\end{eqnarray}

with  $\mathscr{F}^{\mu\nu}_a=\partial^\mu W^\nu_a-\partial^\nu W^\mu_a$ and the $H^a_{\mu\nu\lambda}$ is the field strength of the second rank anti-symmetric field $B^{\mu\nu}_a$, given by
   \begin{eqnarray}
   H^{\mu\nu\lambda}_a&=&\partial^{[\mu}
   B^{\rho\lambda]}_a-gf_{bca}W^{[\mu}_b B^{\rho\lambda]}_c\\ 
   &=&\mathscr{H}^{\mu\nu\lambda}_a-gf_{bca}W^{[\mu}_b
   B^{\rho\lambda]}_c , 
\end{eqnarray}
where $\mathscr{H}^{\mu\nu\lambda}_a= \partial^{[\mu}
B^{\rho\lambda]}_a$.  
  Here the square brackets represent the cyclic permutation of the indices.
  
 The Lagrangian density of eqn(1) is invariant under  $SU(2)$ gauge
 transformations 
  \begin{eqnarray}
  A_\mu\to UA^\mu U^{-1}-\frac{i}{g}\partial^\mu U U^{-1}\qquad
  B^{\mu\nu}\to UB^{\mu\nu}U^{-1} 
  \end{eqnarray}

  The first term in the Lagrangian density (Eq. 8) provides the
  3-point and quartic couplings between the gauge fields. The
  second term term shows $WBB$ and $WWBB$ couplings, where the last
  term, $B\wedge F$, provides a two point derivative coupling
  between $B$ and $W$ and a three-point coupling, $WWB$.

  The presence of the $B\wedge F$ term leads to coupled equations
  of motions in the limit of vanishing gauge coupling $g=0$ 
\begin{eqnarray}
\partial_\mu \mathscr{F}^{\mu\nu}_a=-\frac{m}{6}\epsilon
^{\nu\rho\sigma\lambda}\mathscr{H}_{\rho\sigma\lambda}^a\\ 
\partial_\mu
\mathscr{H}^{\mu\rho\sigma}_a=\frac{m}{2}\epsilon^{\rho\sigma\alpha\beta}
\mathscr{F}_{\alpha\beta}^a 
\end{eqnarray} 
With the gauge-fixing terms 
  \begin{eqnarray}
  \mathscr{L}_{gf}=-\frac{1}{2\xi}(\partial_\mu
  A^\mu_a)^2-\frac{1}{2\zeta}(\partial_\mu B^{\mu\nu}_a)^2 
  \end{eqnarray}
  and with the exclusion of the $B\wedge F$ coupling, the quadratic
  terms in eqn(1) yield the propagators of the $W^a_\mu$ and
  $B^{\rho\lambda}_a$ 
  \begin{eqnarray}
  i\Delta_{\mu\nu,ab} &=&
  -\frac{i}{k^2}\left(g^{\mu\nu}-(1-\xi)\frac{k^\mu
      k^\nu}{k^2}\right) \delta_{ab}\\ 
i\Delta_{\mu\nu,\rho\lambda;ab} &=&\frac{i}{k^2}\left
  (g_{\mu[\rho}g_{\lambda]\nu}-(1-\zeta)\frac{k_{\mu} k_{[\lambda}
    g_{\rho]\nu}-k_{\nu} k_{[\lambda} g_{\rho]\mu}}{k^2}\right)
\delta_{ab} 
  \end{eqnarray}

 The last term in the Lagrangian density gives the vertex rules 
%
%
\begin{eqnarray}
iV_{\mu\nu, \lambda}^{ab} &=& -m \epsilon_{\mu\nu\lambda\rho}k^\rho 
\delta^{ab} \\
iV_{\mu,\nu, \lambda\rho}^{abc} &=& -igm
f^{bca}\epsilon_{\mu\nu\lambda\rho}\,, 
\end{eqnarray}

\begin{figure}[bth]
    \includegraphics[scale=0.4]{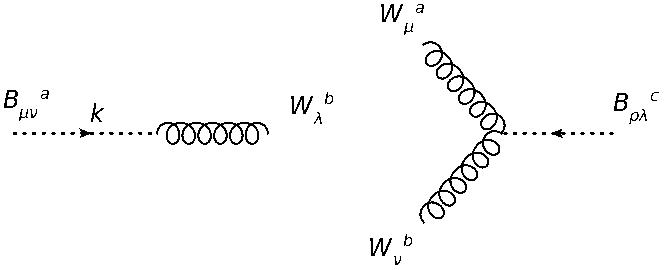} 
\caption{Vertices from the $B\wedge F$ term}
  \label{fig:BF-vertex}
\end{figure}
\begin{figure}[bth]
    \includegraphics[scale=0.4]{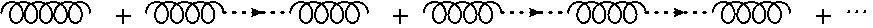} 
\caption{Massive $W$ propagator by summing over $B$ insertions}
  \label{fig:Wprop}
\end{figure}

where the momenta are all directed towards the vertex, and
$\delta^{ab} = 2\Tr({t^at^b})\,.$ In order to use these vertices in
a diagram, we need propagators for the fields, which come from
kinetic terms,
$-\frac{1}{4}\mathscr{F}^a_{\mu\nu}\mathscr{F}^{a\mu\nu}$ for the
$W$ bosons, and
$\frac{1}{12}\mathscr{H}^a_{\mu\nu\lambda}\mathscr{H}^{a\mu\nu\lambda}$
for the $B$ field. In the Feynman-'t~Hooft gauge, and ignoring the
2-point vertex for the moment, the propagators are
\begin{eqnarray}
i\Delta_{\mu\mu'}^{ab} &=& \frac{-ig_{\mu\mu'}}{k^2 +
  i\varepsilon}\delta^{ab} 
\\ 
i\Delta_{\mu\nu, \mu'\nu'}^{ab} &=&
\frac{ig_{\mu[\mu'}g_{\nu']\nu}}{k^2 +
  i\varepsilon}\delta^{ab}\,, 
\label{Bprop}
\end{eqnarray}
where the square brackets indicate antisymmetrization. The
effective tree-level propagator for the $W$ boson is the sum over
all possible insertions of the $B$-field~\cite{Allen:1990gb} as in
Fig.~\ref{fig:Wprop},
\begin{eqnarray}
iD_{\mu\nu} &=& i\Delta_{\mu\nu} +
i\Delta_{\mu\mu'}\frac{1}{2} iV_{\sigma\rho, \mu'} i
\Delta_{\sigma\rho, \sigma'\rho'} \,
\frac{1}{2} iV_{\sigma'\rho', \nu'} i \Delta_{\nu'\nu} + \cdots 
\nonumber \\ 
&=& \frac{-ig_{\mu\nu}}{k^2 + i\varepsilon}\left(1 +
  \frac{m^2}{k^2} + \frac{m^4}{k^4} + \cdots \right) =
\frac{-ig_{\mu\nu}}{k^2 - m^2 + i\varepsilon} \,,
\label{Wprop.massive}
\end{eqnarray}
which is the propagator of a massive vector boson of mass $m$. The
factors of $\frac{1}{2}$ compensate for double-counting due to the
antisymmetrization of the indices.  We have suppressed the gauge
indices here.

~~~~~~~~

The particle interpretation of quantum fields come from the
quadratic part of the Lagrangian.  We have made the vector bosons
massive by diagonalization of the quadratic terms, with no leftover
field degree. The $B$ field has only one degree of freedom per
gauge index, which provides the longitudinal mode of the massive
gauge boson. Thus the $B$ triplet acts similarly to the Goldstone
modes of the complex Higgs field, but the SU(2) symmetry is
unbroken. All the three vector bosons have the same mass, and
nothing analogous to the Higgs particle appears in the spectrum.

\begin{figure}[bth]
    \includegraphics[scale=0.4]{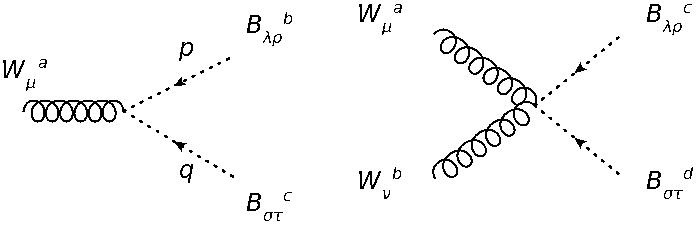} 
\caption{Vertices from the $H_{\mu\nu\lambda}H^{\mu\nu\lambda}$ term }
  \label{fig:BBA-BBAA}
\end{figure}

\begin{figure}[bth]
    \includegraphics[scale=0.4]{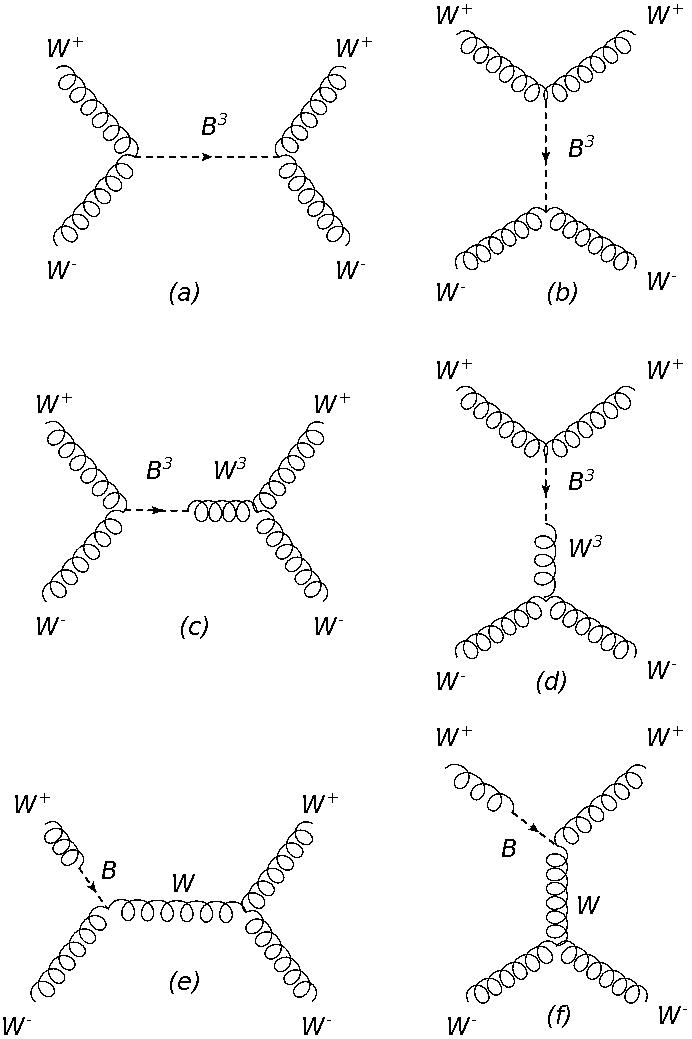} 
\caption{Scattering diagrams with $P^2$ behavior: I}
  \label{fig:one-B}
\end{figure}

\begin{figure}[bth]
    \includegraphics[scale=0.4]{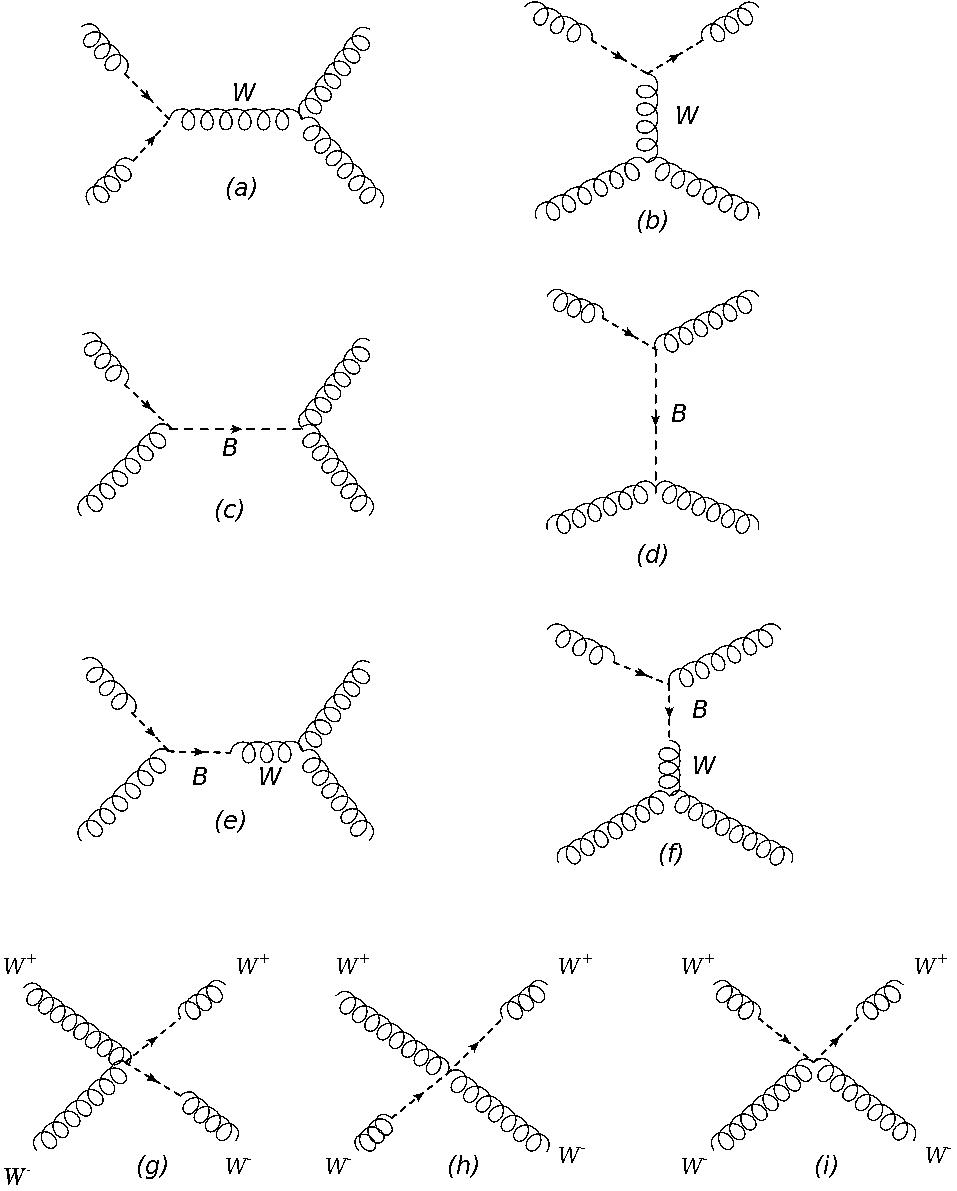} 
\caption{Scattering diagrams with $P^2$ behavior: II}
  \label{fig:two-B}
\end{figure}

\begin{figure}[bth]
    \includegraphics[scale=0.4]{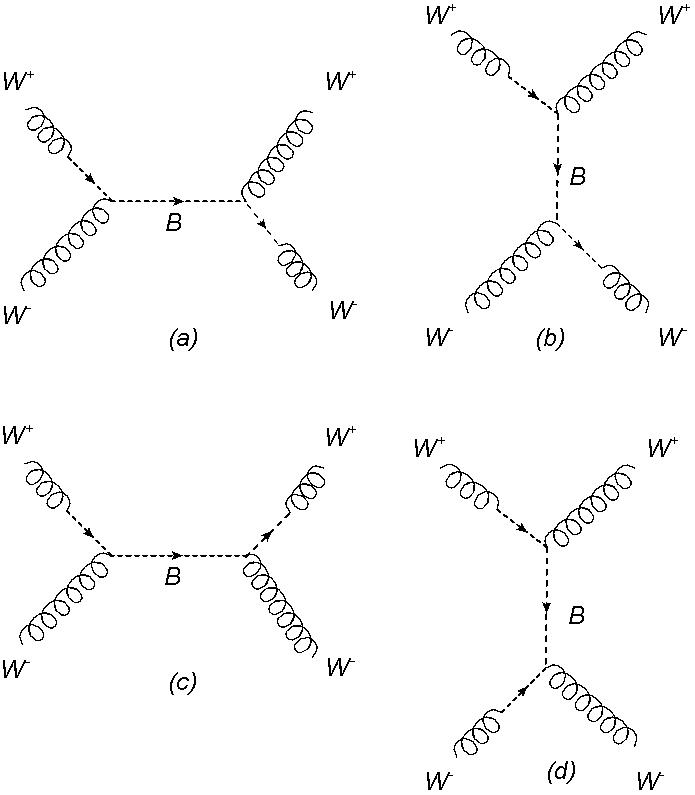} 
\caption{Scattering diagrams with $P^2$ behavior: III}
  \label{fig:three-B}
\end{figure}

However, the kinetic term for the $B$ field contains new
interactions between the $B$ and the $W$ fields. From eq(1) and
using the eq(11), we can read off the vertices shown in
Fig.~\ref{fig:BBA-BBAA},
\begin{eqnarray}
iV^{abc}_{\mu, \lambda\rho, \sigma\tau} &=& gf^{abc}\left[(p - q)_\mu
g_{\lambda[\sigma}g_{\tau]\rho} +
p_{[\sigma}g_{\tau][\lambda}g_{\rho]\mu} -
q_{[\lambda}g_{\rho][\sigma}g_{\tau]\mu}  \right] \\
iV^{abcd}_{\mu, \nu, \lambda\rho, \sigma\tau} &=& ig^2 \left[
  f^{ace}f^{bde} 
\left(g_{\mu\nu}g_{\lambda[\sigma}g_{\tau]\rho} +
  g_{\mu[\sigma}g_{\tau][\lambda}g_{\rho]\nu} \right) 
+ f^{ade}f^{bce}\left(g_{\mu\nu}g_{\lambda[\sigma}g_{\tau]\rho} +
  g_{\mu[\lambda}g_{\rho][\sigma}g_{\tau]\nu}\right) 
     \right]\,.
\end{eqnarray}
Instead of the diagrams of Fig.~\ref{fig:H-mediated} mediated by
the Higgs particle we now find several new diagrams corresponding
to $W^+ W^-\to W^+ W^-$ scattering at the tree level. We may ignore
those which, by power counting, behave as $P^0$ or less.  We have
grouped the remaining diagrams into Fig.~\ref{fig:one-B},
Fig.~\ref{fig:two-B} and Fig.~\ref{fig:three-B}, according to the
number of internal $B$ propagators. The amplitudes for all of these
diagrams go as $P^2$.
In Fig.~\ref{fig:one-B}, diagrams (a) and (b) appear only once, but
diagrams (c) and (d) have twins, obtained by exchanging the
internal $B$ and $W$ lines. Similarly, the $B$ line can be on any
of the external legs in each of diagrams (e) and (f), leading to a
multiplicity of 4.

We have calculated the amplitudes corresponding to these diagrams
using the vertex rules and propagators given above. For the
internal $W$ propagators we have used the resummed propagator of
Eq.~(\ref{Wprop.massive}). For internal $B$ propagators we can do a
similar resummation, leading again to $k^2 - m^2$ in the
denominator. For a $B$ propagator on an external leg, we have used
the propagator in Eq.~(\ref{Bprop}). 
The amplitudes for the diagrams in Fig.~\ref{fig:one-B}, including
their multiplicities, are
\begin{eqnarray}
{\cal M}_{\ref{fig:one-B}a} + {\cal M}_{\ref{fig:one-B}b} &=&
-\frac{3g^2 P^2}{2m^2} (1 + c)  + O(P^0)\\ 
2\left({\cal M}_{\ref{fig:one-B}c} + {\cal
 M}_{\ref{fig:one-B}d}\right) &=& \frac{3g^2 P^2}{m^2} (1 + c)  +
O(P^0) \\ 
4\left({\cal M}_{\ref{fig:one-B}e} + {\cal
 M}_{\ref{fig:one-B}f}\right) &=& -\frac{2g^2 P^2}{m^2} (1 + c)  +
O(P^0) \,.
\end{eqnarray}
We note that the diagrams in the last row of Fig.~\ref{fig:two-B}
have different amplitudes, even though the diagrams themselves
appear to be related by exchanges of $B$ and $W$ lines. Including
multiplicities, the amplitudes of Fig.~\ref{fig:two-B} are
\begin{eqnarray}
2\left({\cal M}_{\ref{fig:two-B}a} + {\cal
    M}_{\ref{fig:two-B}b}\right)  &=& \frac{2g^2 P^2}{m^2} (1 + c)
+ O(P^0) \\
4\left({\cal M}_{\ref{fig:two-B}c} + {\cal M}_{\ref{fig:two-B}d}
\right) &=& \frac{4g^2 P^2}{m^2} (1 + c) + O(P^0) \\ 
4\left({\cal M}_{\ref{fig:two-B}e} + {\cal M}_{\ref{fig:two-B}f}
\right) &=& -\frac{4g^2 P^2}{m^2} (1 + c) + O(P^0) \\ 
2\left({\cal M}_{\ref{fig:two-B}g} + {\cal M}_{\ref{fig:two-B}h} +
  {\cal   M}_{\ref{fig:two-B}i}\right) &=& \frac{2g^2 P^2}{m^2} 
(1 + 3c + 2c^2)  + O(P^0)\,.
\end{eqnarray}
The remaining diagrams which go as $P^2$ are shown in
Fig.~\ref{fig:three-B}. There are two of each diagram,
corresponding to exchanging the $B$ line between the 
incoming lines and  simultaneously between the outgoing lines.
The amplitudes for these are
\begin{eqnarray}
2\left({\cal M}_{\ref{fig:three-B}a} + {\cal
    M}_{\ref{fig:three-B}b} + {\cal M}_{\ref{fig:three-B}c} 
    + {\cal M}_{\ref{fig:three-B}d}\right) = 
    -\frac{4g^2 P^2}{m^2} (1 + 2c
+c^2)  + O(P^0)\,.
\end{eqnarray}
Adding the amplitudes of the diagrams in Fig.~\ref{fig:one-B},
Fig.~\ref{fig:two-B} and Fig.~\ref{fig:three-B}, we get 
\begin{eqnarray}
{\cal M}_6 + {\cal M}_7 + {\cal M}_8 =  -\frac{g^2 P^2}{2m^2} (1 +
c) + O(P^0)\,,
\end{eqnarray}
which, when added to the amplitudes of the purely $W$-mediated
diagrams of Fig.~\ref{fig:Wonly}, cancels the $P^2$ divergence
exactly. The $W_L^+ W_L^-$ elastic scattering amplitude remains
finite as $P\to \infty\,$, and unitarity is not violated. We note
that there are other diagrams in this model for the  $W_L^+ W_L^-$
elastic scattering process, but all those are of the order $P^0$,
so do not affect our argument. 

The processes, $W^+_L W^+_L\to W^+_L W ^+_L$ , $W^-_L W^-_L\to
W^-_L W^-_L$ and $W^+_L W^-_L\to W^3_L W^3_L$ can be shown in a
similar manner to be finite at high energy.

This demonstration of unitarity should not come as a great
surprise.  The action can be written in a form invariant under the
Becchi-Rouet-Stora-Tyutin (BRST)-symmetry by using the auxiliary
field in a Stueckelberg mechanism~\cite{Ruegg:2003ps} for the
$B$-field, which shows that the theory is
unitary~\cite{Lahiri:1996dm}.

It can also be shown~\cite{Lahiri:1999uc} that perturbative
corrections contribute no new terms to the action, but only to a
rescaling, or renormalization, of the couplings and masses.

We conclude that the topological, or $B\wedge F$, mass generation
mechanism for 4-dimension SU(2) gauge theory with the Lagrangian as
given in Eq(8) does not violate tree level unitarity of scattering
amplitude of longitudinal gauge bosons.



\end{document}